\documentclass[a4paper]{jpconf}
\usepackage{graphicx}
\begin{document}
\title{Multiple frustration-induced plateaus in a magnetization process
       of the mixed spin-$1/2$ \\ and spin-$3/2$ Ising-Heisenberg diamond chain}

\author{J Stre\v{c}ka, L \v{C}anov\'a, T Lu\v{c}ivjansk\'y, and M Ja\v{s}\v{c}ur}

\address{Department of Theoretical Physics and Astrophysics, Faculty of Science, \\ 
P. J. \v{S}af\'{a}rik University, Park Angelinum 9, 040 01 Ko\v{s}ice, Slovak Republic}
\ead{lucia.canova@upjs.sk}

\begin{abstract}
Magnetization process of the mixed spin-$1/2$ and spin-$3/2$ Ising-Heisenberg diamond chain is 
examined by combining three exact analytical techniques: Kambe projection method, decoration-iteration transformation and transfer-matrix method. Multiple frustration-induced plateaus in a magnetization process of this geometrically frustrated system are found provided that a relative ratio between the antiferromagnetic Heisenberg- and Ising-type interactions exceeds some particular value. By contrast, there is just a single magnetization plateau if the frustrating Heisenberg interaction is sufficiently small compared to the Ising one.
\end{abstract}

\section{Introduction}

The spin-1/2 quantum Heisenberg model on the diamond chain has enjoyed a considerable research 
interest since two unusual tetramer-dimer and dimer-monomer phases were theoretically predicted 
by Takano \textit{et al}. \cite{Tak96} as a consequence of the mutual interplay between quantum fluctuations and geometric frustration. Another stimulus for investigating diamond chain models 
was triggered by a recent discovery of molecular solids ${\rm Bi_4Cu_3V_2O_{14}}$ \cite{Sak02}, 
${\rm Cu}_3({\rm CO}_3)_2({\rm OH})_2$ \cite{Kik05}, ${\rm Cu_3(TeO_3)_2Br_2}$ \cite{Uem07}, 
which represent possible experimental realizations of the diamond chain compounds. Motivated 
by these experiments, several phenomenological models have been suggested and solved with 
the aim to bring insight into a frustrated magnetism of the diamond chain compounds 
(see \cite{Can06} and references therein). In the present work, we will develop an exact 
analytical treatment for the mixed spin-$1/2$ and spin-$3/2$ Ising-Heisenberg diamond chain 
that should clarify how the geometric frustration influences a possible magnetization scenario.

\section{Model and its exact solution}

Let us write the total Hamiltonian of the mixed spin-$1/2$ and spin-$3/2$ Ising-Heisenberg diamond chain as a sum over plaquette Hamiltonians $\hat{\mathcal H} = \sum_k \hat{\mathcal H}_k$, where each plaquette Hamiltonian $\hat{\mathcal H}_k$ involves all the interaction terms included within $k$th diamond-shaped plaquette
\begin{eqnarray}
\hat{\mathcal H}_k = J_{\rm H} \vec{\bf S}_{3k-1} \!\!\cdot\! \vec{\bf S}_{3k}
                  \!+\! J_{\rm I} ({\hat{S}^{z}}_{3k-1} \!+\! {\hat{S}^{z}}_{3k})   
                              ({\hat{\mu}^{z}}_{3k-2} \!+\! {\hat{\mu}^{z}}_{3k+1}) 
                  \!-\! H_{\rm H} ({\hat{S}^{z}}_{3k-1} \!+\! {\hat{S}^{z}}_{3k}) 
                  \!-\! H_{\rm I} ({\hat{\mu}^{z}}_{3k-2} \!+\! {\hat{\mu}^{z}}_{3k+1})/2.
\label{eq1}
\end{eqnarray}
Here, ${\hat{\mu}^{z}}_{k}$ and $\vec{\bf S}_{k} = ({\hat{S}^{x}}_{k}, {\hat{S}^{y}}_{k}, {\hat{S}^{z}}_{k})$ denote spatial components of the spin-$1/2$ and spin-$3/2$ operators,
the parameter $J_{\rm H}>0$ labels the isotropic antiferromagnetic interaction between 
the nearest-neighbouring Heisenberg spins and the parameter $J_{\rm I}>0$ accounts for the antiferromagnetic Ising-type interaction between the Heisenberg spins and their nearest Ising neighbours. The last two terms determine the magnetostatic Zeeman's energy of the Ising and Heisenberg spins in an external magnetic field $H_{\rm I}$ and $H_{\rm H}$ oriented along the $z$-axis.

Exact solution for the aforedescribed Ising-Heisenberg diamond chain can be obtained by combining three exact analytical techniques: the Kambe projection method \cite{Kam50}, the generalized decoration-iteration transformation \cite{Fis59} and the transfer-matrix method \cite{Bax82}. First, let us take advantage of a commutation relation between different plaquette Hamiltonians $[\hat{\mathcal H}_i, \hat{\mathcal H}_j] = 0$ valid for each $i \neq j$ to partially factorize
the partition function 
\begin{eqnarray}
{\mathcal Z} = \sum_{\{ \mu_k \}} \Tr_{\{ S_k \}} \exp(-\beta\hat{{\mathcal H}})
     = \sum_{\{ \mu_k \}} \prod_{k=1}^N \Tr_{k} \exp(-\beta\hat{{\mathcal H}_k}).
\label{eq2}	
\end{eqnarray}
Here, $\beta = 1/(k_{\rm B} T)$ ($k_{\rm B}$ is Boltzmann's constant, $T$ is the absolute temperature), the symbols $\sum_{\{ \mu_k \}}$ and $\Tr_{\{ S_k \}}$ denote a summation over spin degrees of freedom of all Ising and Heisenberg spins, respectively. To proceed further with calculation, it is necessary to perform a trace over spin degrees of freedom of two Heisenberg spins from $k$th diamond plaquette as indicated by the symbol $\Tr_{k}$ shown on the rhs of Eq.~(\ref{eq2}). This calculation can easily be accomplished by making use of the Kambe projection method \cite{Kam50}, since the Hamiltonian $\hat{\mathcal H}_k$ can also be viewed as the Hamiltonian of 
the spin-$3/2$ Heisenberg dimer in the effective field
$H_{\rm eff} = H_{\rm H} - J_{\rm I} (\mu^{z}_{3k-2} + \mu^{z}_{3k+1})$. Consequently, 
the complete set of eigenvalues $E_k$ corresponding to the plaquette Hamiltonian $\hat{\mathcal H}_k$ can be expressed solely as a function of two quantum spin numbers $S_{\rm tot}$ and $S_{\rm tot}^z$,
\begin{eqnarray}
E_k (S_{\rm tot}, S_{\rm tot}^z) = - 15 J_{\rm H}/4 + J_{\rm H} S_{\rm tot} (S_{\rm tot} + 1)/2 
                               - S_{\rm tot}^z H_{\rm eff} 
                               - H_{\rm I} (\mu^{z}_{3k-2} + \mu^{z}_{3k+1})/2,
\label{eq3}	
\end{eqnarray}
which determine the total quantum spin number of the spin-$3/2$ Heisenberg dimer and its 
projection towards the $z$-axis. According to the basic laws of quantum mechanics, the total 
quantum spin number of the spin-3/2 dimer gains four different values 
$S_{\rm tot} = 0, 1, 2, 3$, while the quantum spin number $S_{\rm tot}^z$ gains  $2S_{\rm tot}+1$ different values $S_{\rm tot}^z = -S_{\rm tot}, -S_{\rm tot}+1, \ldots, S_{\rm tot}$ for each allowed value of $S_{\rm tot}$. The energy eigenvalues (\ref{eq3}) can be then used to obtain the relevant trace of the last expression on rhs of Eq.~(\ref{eq2}), whose explicit form immediately implies a possibility of performing the generalized decoration-iteration transformation \cite{Fis59}
\begin{eqnarray}
\Tr_{k} \exp(-\beta\hat{{\mathcal H}_k})
&=& \exp[\beta H_{\rm I}(\mu^{z}_{3k-2} + \mu^{z}_{3k+1})/2] 
\sum_{n=0}^{3} \sum_{m=-n}^{n} \! \!  
\exp [-\beta J_{\rm H}n(n + 1)/2]\cosh(\beta H_{\rm eff} m)
\nonumber \\
&=& A \exp[\beta R \mu^{z}_{3k-2} \mu^{z}_{3k+1} 
          + \beta H (\mu^{z}_{3k-2} + \mu^{z}_{3k+1})/2].
\label{eq4}	
\end{eqnarray}
Note that unimportant constant term $\exp(15 \beta J_{\rm H}/4)$ has been for simplicity 
omitted from lhs of Eq.~(\ref{eq4}) as it does not basically change physics of the studied 
system. It should be mentioned, however, that the decoration-iteration transformation (\ref{eq4}) 
must hold for arbitrary combination of spins states of two Ising spins $\mu_{3k-2}$ and $\mu_{3k+1}$ involved therein. This self-consistency condition unambiguously determines so far not specified 
mapping parameters $A$, $R$ and $H$
\begin{eqnarray}
A = (W_+W_-W^2)^{1/4}, \quad 
\beta R = \ln (W_{+}W_{-}) - 2\ln W,
\quad \beta H = \beta H_{\rm I} - \ln W_{+} + \ln W_{-},
\label{eq5}	
\end{eqnarray}
which can be uniquely expressed through the functions $W_{\pm} = F(\pm 1)$ and $W = F(0)$, where
\begin{eqnarray}
F(x) = \sum_{n=0}^{3} \sum_{m=-n}^{n} \! \!
 \exp [-\beta J_{\rm H}n(n + 1)/2]
\cosh[\beta m (J_{\rm I} x + H_{\rm H})].
\label{eq6}	
\end{eqnarray}
It is easy to verify that the transformation (\ref{eq4}) establishes a precise mapping relationship  between the mixed spin-1/2 and spin-$3/2$ Ising-Heisenberg diamond chain and its corresponding spin-1/2 Ising linear chain with the nearest-neighbour coupling $R$ and the effective magnetic field $H$ given by Eqs.~(\ref{eq5}) and (\ref{eq6}). A direct substitution of the mapping transformation (\ref{eq4}) into Eq.~(\ref{eq2}) actually connects partition functions of both these models through the relation
\begin{eqnarray}
{\mathcal Z} (\beta, J_{\rm H}, J_{\rm I}, H_{\rm H}, H_{\rm I}) 
            = A^N {\mathcal Z}_{\rm Ising} (\beta, R, H).
\label{eq7}	
\end{eqnarray}
It is noteworthy that the partition function ${\mathcal Z}_{\rm Ising}$ of the spin-1/2 Ising linear chain can simply be attained with the help of the transfer-matrix method (see for instance Ref.~\cite{Bax82}) and thus, our exact calculation is formally completed. In this respect, 
the mapping relation (\ref{eq7}) can also be used to obtain exact results for the Helmholtz 
free energy ${\mathcal F} = - \beta^{-1} \ln {\mathcal Z}$ and subsequently, both the sublattice magnetization $m_{\rm H} = - \frac{1}{2N} \left( \frac{\partial {\mathcal F}}{\partial H_{\rm H}} \right)$ and $m_{\rm I} = - \frac{1}{N} \left( \frac{\partial {\mathcal F}}{\partial H_{\rm I}} \right)$.

\section{Results and discussion}

Now, let us step forward to a discussion of the most interesting numerical results obtained 
for magnetization process. Before doing so, it is useful to mention that magnetic properties 
of the frustrated Ising-Heisenberg diamond chain with other two particular values of the 
Heisenberg spins $S = 1/2$ and $1$ have been detailed discussed in our previous work~\cite{Can06}. 
The main objective of the present work is therefore to investigate magnetization curves of 
the mixed spin-$1/2$ and spin-$3/2$ Ising-Heisenberg diamond chain, which should serve for 
the sake of comparison with the recently published results \cite{Can06} and would shed light 
on how the scenario of magnetization process depends on a magnitude of the Heisenberg spins.

\begin{figure}
\vspace*{-0.6cm}
\begin{center}
\hspace*{-0.8cm}
\includegraphics[width=17.5cm]{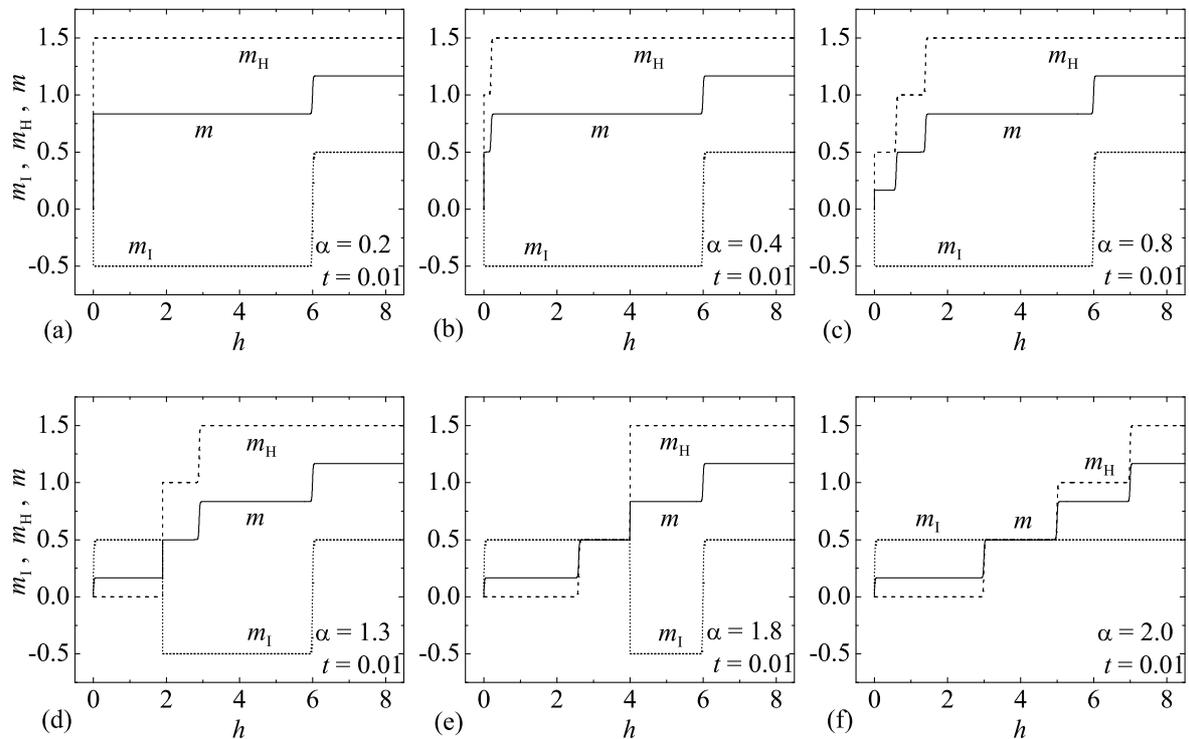}
\vspace*{-1.0cm}
\caption{The total magnetization $m$ (solid lines) and the sublattice magnetization $m_{\rm I}$ (dotted lines), $m_{\rm H}$ (dashed lines) as functions of the external magnetic field
at very low temperature $t = 0.01$ for several values of the frustration parameter 
(a)~$\alpha = 0.2$, (b)~$\alpha = 0.4$, (c)~$\alpha = 0.8$, (d)~$\alpha = 1.3$, 
(a)~$\alpha = 1.8$, (b)~$\alpha = 2.0$.}
\label{fig:mH}
\end{center}
\end{figure}

To simplify further discussion, let us first rescale all interaction parameters with respect 
to the Ising-type interaction $J_{\rm I}$ by introducing the following set of dimensionless parameters: 
$\alpha = J_{\rm H}/J_{\rm I}$, $h = H_{\rm I}/J_{\rm I} = H_{\rm H}/J_{\rm I}$ and $t = k_{\rm B}T/J_{\rm I}$ as describing a strength of the geometric frustration, external field and temperature, respectively. The total magnetization [$m=(m_{\rm I} + 2 m_{\rm H})/3$] together with both sublattice magnetization $m_{\rm I}$ and $m_{\rm H}$ is plotted in Fig.~\ref{fig:mH} against the reduced 
magnetic field for several values of the frustration parameter $\alpha$. As one can see, we have 
altogether found six different types of magnetization curves. If the frustration parameter is 
$\alpha < 1/3$ [see Fig.~\ref{fig:mH}(a)], there exists just
a single intermediate magnetization plateau before the magnetization tends towards its saturation value. On the other hand, two or three intermediate magnetization plateaus can be observed provided that 
the frustration parameter is from the interval $1/3 < \alpha < 1/2$ [Fig.~\ref{fig:mH}(b)] or $\alpha > 1/2$ [see Figs.~\ref{fig:mH}(c)--(f)], respectively.
It is quite obvious from this analysis that the geometric frustration can be regarded
as the driving force behind appearance of the greather number of intermediate magnetization plateaus.

It is worthwhile to remark that the identified magnetization plateaus appear at $1/7$, $3/7$ and $5/7$ of the saturation magnetization when normalizing the total magnetization with respect to its saturation value. All these fractional values are in accordance with the Oshikawa-Yamanaka-Affeck rule~\cite{Osh97}, which has been proposed as a necessary condition for the formation of quantized plateaus. As it could be readily 
understood from the relevant results for the sublattice magnetization, the spin arrangements 
emerging at intermediate plateau regions with the same value of the total magnetization 
might significantly differ. However, the precise nature of spin arrangements to emerge 
at different plateau regions requires a more comprehensive analysis of the ground-state 
phase diagram, which is left as a challenging task for our future work.

It should be also mentioned that the stepwise magnetization curves with an abrupt change(s)
of the magnetization observable at critical fields occur merely at zero temperature, while 
there are no real magnetization jumps at any finite temperatures. It turns out, however, that 
the magnetization curve exhibits a rather rapid continuous increase in the vicinity of critical 
fields whenever the temperature is set sufficiently small. Contrary to this, the magnetization 
curves are gradually smeared out when the temperature is raised from zero and consequently, 
the plateau structure gradually diminishes from the magnetization curves above $t \sim 0.4$.

In conclusion, we have developed an exact analytical treatment that enables to study the magnetization process of the mixed spin-$1/2$ and spin-$3/2$ Ising-Heisenberg diamond chain. Within the framework of this exact method, we have found a rigorous evidence for multiple frustration-induced plateaus, whereas the number of intermediate plateaus basically depends on a strength of the frustration parameter $\alpha$. It is noteworthy, moreover, that the developed exact analytical treatment can be rather easily generalized to account for the magnetization scenario of the mixed spin-$1/2$ and spin-$S$ Ising-Heisenberg diamond chain, which will be dealt with in our subsequent more 
comprehensive study.

\ack{This work was supported by the Slovak Research and Development Agency under the contract 
LPP-0107-06 and by Ministry of Education of SR under the grant No.~VEGA~1/0128/08.}

\end{document}